\documentclass[preprint,12pt]{elsarticle}

\usepackage{amssymb}
\usepackage{amsmath}
\usepackage{graphicx}
\usepackage{framed,multirow}
\usepackage{url}

\journal{Information fusion}

\begin{document}

\begin{frontmatter}



\title{ICFNet: Integrated Cross-modal Fusion Network for Survival Prediction}

\author[1]{Binyu Zhang \corref{co-first}}
\author[1]{Zhu Meng \corref{co-first}}
\author[1]{Junhao Dong}
\author[1]{Shichao Li}
\author[1,2]{Fei Su}
\author[1,2]{Zhicheng Zhao \corref{cor1}}
\cortext[co-first]{These authors contribute equally.}
\cortext[cor1]{Corresponding author: E-mail: zhaozc@bupt.edu.cn.}

\address[1]{School of Artificial Intelligence, Beijing University of Posts and Telecommunications, Beijing, 100876, China}
\address[2]{Beijing Key Laboratory of Network System and Network Culture}

\begin{abstract}
Survival prediction is a crucial task in the medical field and is essential for optimizing treatment options and resource allocation. However, current methods often rely on limited data modalities, resulting in suboptimal performance. In this paper, we propose an Integrated Cross-modal Fusion Network (ICFNet) that integrates histopathology whole slide images, genomic expression profiles, patient demographics, and treatment protocols. Specifically, three types of encoders, a residual orthogonal decomposition module and a unification fusion module are employed to merge multi-modal features to enhance prediction accuracy. Additionally, a balanced negative log-likelihood loss function is designed to ensure fair training across different patients. Extensive experiments demonstrate that our ICFNet outperforms state-of-the-art algorithms on five public TCGA datasets, including BLCA, BRCA, GBMLGG, LUAD, and UCEC, and shows its potential to support clinical decision-making and advance precision medicine. The codes are available at: \url{https://github.com/binging512/ICFNet}.
\end{abstract}


\end{frontmatter}



\section{Introduction}
\label{sec:introduction}
Recently, deep learning has exhibited great potential for various medical tasks \cite{zhang2024mtcsnet, meng2024nusea}. In the field of cancer research, survival analysis plays a vital role in estimating the death risk of patients in the cancer prognosis. By scrutinizing the survival durations of patients, researchers endeavor to pinpoint key determinants and assess the efficacy of therapeutic interventions. 

An increasing number of deep learning-based methods for survival prediction have been proposed \cite{lu2021data}. As the gold standard for diagnosing cancers, histopathology whole slide images (WSIs) are widely utilized for survival analysis. Due to the extreme scale of WSIs and the limitation of graphics processing units (GPU) computing power, multi-instance learning (MIL) is usually adopted for WSI-based methods. In parallel, the performance of the Transformer model \cite{vaswani2017attention} and CLIP \cite{dosovitskiy2020image} in image analysis opens up a path to use attention-based models for survival prediction. 

However, although WSIs can provide the morphological information of tumor tissue and the hints for cancer diagnosis, the inherent sparse distribution of vital information within WSIs and the training methodology of MIL lead to sub-optimal model performance. Hence, some researches consider to utilize the genomics data to improve the accuracy of survival prediction, because cancer patients exhibit intrinsic genetic mutations. By conducting genomics analysis of patients, a more precise understanding of their cancer condition can be achieved \cite{klambauer2017self}. As studies have shown that the genes expression cam reveal the morphological characteristics in histopathology WSI \cite{lin2019ghrelin,wang2019mmp}, some methods \cite{chen2021multimodal,xu2023multimodal} attempt to use both histopathology and genomics data. They leverage multi-modal network to extract and fuse the features from different modality by calculating the similarity score between each histo-genomics data pair. However, histopathology and genomics data are still not able to provide enough information for prognosis. In fact, the survival of patients depends not only on their clinical diagnosis, but also on variety of other factors, such as their physical conditions, demographic characteristics and the treatment they receive.

Therefore, aiming to achieve more precise prognostic predictions with various information we propose an Integrated Cross-modal Fusion Network (ICFNet) to incorporate the demographic data and the treatment protocols. ICFNet takes four types of data as input, including histopathology WSIs, genomics data, patient information and treatment protocols. As the input data is of different modalities, ICFNet employs different strategies to extract the features. Specifically, following \cite{chen2021multimodal}, the histopathology WSIs are split into patches and encoded with pre-trained image models, while genomics data are divided into groups to extract the features. As for the patient information and the treatment protocols, two types of text prompts are designed, and a pre-trained language model is utilized to encode the text. As three modalities of data are acquired in ICFNet, optimal-transport-based co-attention Transformers are leveraged to extract the relevant features among the modalities. To reduce redundancy among modality and enhance modality-specific features, ICFNet introduces a residual orthogonal decomposition (ROD) module to improve information utilization. Due to the variation of features in different modalities, ICFNet employs a unification fusion module to align the features of different modalities into a shared latent space, then a densely supervised scheme is applied to each modality to enhance the learning capability of the network. In addition, during survival prediction, traditional Negative Log-Likelihood Loss (NLLLoss)-based \cite{zadeh2020bias} cumulative probability operation will result in labels imbalanced. To address this issue, we propose a balanced NLLLoss function.

Furthermore, the proposed ICFNet can be considered as an auxiliary decision-making tool for therapeutic approaches in future clinical practice. By inputting the patients' laboratory test results, demographic, and planned treatment protocols into the network, the network can evaluate the effectiveness of the treatment options by the estimated risk. This evaluation can further assist healthcare experts in making better treatment decisions to avoid over-treatment and wastage of healthcare resources.

Our contributions can be summarized as follows:
\begin{itemize}
\item A multi-modal survival prediction framework named ICFNet is constructed via MIL fashion, and four types of patient data including histopathology WSIs, genomic expression data, demographic information, and treatment protocols are incorporated. To address the unique characteristics of each data type, three different modal encoders are utilized to extract distinct features among each modality and optimal-transport (OT) algorithm are employed to extract interrelated features among modalities.
\item To reduce redundancy among modality and enhance modality-specific features, ROD module is introduced. To transform features from diverse modalities into a common latent space, ICFNet incorporates a unification fusion module. Additionally, a Balanced NLLLoss is presented to mitigate the issue of imbalanced loss computation across different labels, enhancing the model's performance.
\item Extensive experiments on five public datasets in The Cancer Genome Atlas (TCGA), namely BLCA, BRCA, GBMLGG, LUAD and UCEC, demonstrate that ICFNet achieves the state-of-the-art performance for survival prediction task.
\end{itemize}

The remainder of the paper proceeds as follows: Section \ref{sec:related works} introduces the related work on survival prediction. Section \ref{sec:methodology} is concerned with the specific methodology of ICFNet. Section \ref{sec:experiments} analyzes the results of experiments on the five public datasets. Section \ref{sec:conclusion} provides the conclusions.

\section{Related Works}
\label{sec:related works}
\subsection{Multi-instance learning for medical image analysis}
Due to the inherent gigapixel nature of WSIs, MIL has been widely adopted as a bag-level supervised approach. Impressive results had been achieved. For example, AttnMIL \cite{ilse2018attention} utilized the attention mechanism to aggregate patch features, while Raju et al. \cite{raju2020graph} first constructed the patch instances into graphs with distance and cluster information, then proposed a graph attention network to fuse patch features for accurate cancer staging. To overcome the challenge of training patch feature extractors, C2C \cite{sharma2021cluster} proposed an end-to-end framework. DSMIL \cite{li2021dual} introduced a self-supervised contrastive learning method with a masked non-local block designed to aggregate the instances. TransMIL \cite{shao2021transmil} co-related Transformer \cite{vaswani2017attention} and MIL into the framework. To explore the enlarged patch bags in MIL, DTFD-MIL\cite{zhang2022dtfd} introduced the concept of pseudo-bags and proposed a double-tier MIL framework to effectively use intrinsic features. To relieve the overfitting problem caused by limited samples, SH-Transformers \cite{yan2023sparse} introduced a sparse attention mechanism to learn the hierarchical WSI representation. Attempting to find the critical information among thousands of patches, dMIL-Transformer \cite{chen2023dmil} proposed a double Max-Min MIL strategy to select the suspected top-K positive patches to further improve the inference. Considering different magnifications of WSIs were likely to have positive contribution for diagnosis, TGMIL \cite{sun2023tgmil}, CS-MIL \cite{deng2024cross} and Tsiknakis et al. \cite{tsiknakis2024unveiling} incorporated three different magnifications and combined features with different mechanisms. To investigate the interpretability, SI-MIL \cite{kapse2024si} provided an interpretable-by-design MIL method. These impressive approaches highlighted the effectiveness of MIL in analyzing WSIs and consequently were facilitated to various downstream medical tasks, including survival prediction.

\subsection{Cancer survival prognosis}
Prognosis, as one of the most critical tasks in clinical practice, aims to predict patient survival duration and prevent over-treatment or wastage of medical resources as well. Consequently, numerous deep learning-based approaches for patient prognosis have been proposed. DeepAttnMISL \cite{yao2020whole} introduced both siamese MI-FCN and attention-based MIL pooling to efficiently learn imaging features and then aggregated WSI-level information to patient-level. Aiming to better aggregate instance-level histology features to model local- and global-level topological structures, Patch-GCN \cite{chen2021whole} treated WSIs as 2D point clouds and leveraged GCN for feature fusion. Sandarenu et al. \cite{sandarenu2022survival} attempted to cluster the patch features and fuse them with attention modules. Surformer \cite{wang2023surformer} was proposed to quantify specific histological patterns via bag-level labels. As a distinctive characteristic of cancer patients, their gene expression undergo alterations. Consequently, SNN \cite{klambauer2017self} employed genomics information to predict patient survival. To give model more supervision and enhance the model performance, LNPL-MIL\cite{shao2023lnpl} and HistMIMT \cite{shao2024multi} introduced multi-tasks, such as diagnosis, genomics expression prediction and survival prediction. However, these methods typically consider information from only a single modality or limited data, resulting in a partial understanding of the patient, consequently, prognostic models could only achieve sub-optimal results.

To gain a more comprehensive understanding of patient conditions, several existing approaches took multi-modal data into account, and leveraged diverse information sources to integrally predict patient survival time. As the WSI and genomics data are critical for cancer diagnosis, Pathomic \cite{chen2020pathomic} encoded WSI patches, cell spatial graphs and genomic profiles and fused the features for survival prediction. However, there was no interaction between the extracted features. Therefore, MCF\cite{he2020feasibility} proposed a hierarchical predictive scheme to reliably link the multi-modality features and multiple classifiers, while MCAT \cite{chen2021multimodal} proposed a genomic-guided co-attention module to assist the network in learning genomics-related features from WSIs. Although the co-attention improved the model performance, it focused on only dense local similarity across modalities, thereby failed to capture global consistency between potential structures. Thus, MOTCat \cite{xu2023multimodal} utilized the OT mechanism to overcome the problem. To fully leverage the relationships between various modalities, SurvPath \cite{jaume2024modeling} introduced a multi-modal dense interaction network, facilitating the fusion of information from different modalities at multiple levels, while CMTA\cite{zhou2023cross} proposed a framework to explore the intrinsic cross-modal correlations and transfer potential complementary information. To learn the prototype for survival prediction, PIBD\cite{Zhang2024PROTOTYPICAL} introduced two kinds of module to reduce the intra- and inter-modal redundancy, while MMP\cite{Song2024Multimodal} constructed an unsupervised and compact WSI representation with a Gaussian mixture model. To decpmpose multi-modal knowledge into distinct components, CCL\cite{zhou2024cohort} introduced an MKD module and a CGM scheme to train the network. To reduce the computational cost of the model in practical applications, G-HANet \cite{wang2024histo} proposed a framework with knowledge distillation. It reconstructed gene features from histopathology image features, aiming to minimize the need for gene expression data during testing. Moreover, some methods adopted other medical data for prognosis. HMCAT \cite{li2023survival} and Jeong et al.\cite{jeong2024artificial} proposed multi-modal frameworks to explore the relationships among other modalities, such as histopathology WSIs, computed tomography (CT) images and magnetic resonance images (MRI).

Although these approaches attempted to consider multi-modal data and trained with well-designed models, it was still not sufficient to predict the patients survival with only one or two types of data. Moreover, they didn't explore patient demographics and treatment information, which would substantially influence patient survival time.

\section{Methodology}
\label{sec:methodology}

\subsection{Overview and problem formulation}
MIL strategy is also adopted for the network training and data bags are formulated in advance.

\textbf{WSI bags formulation.} As the tissue in WSI is likely to be sparsely distributed, a foreground segmentation algorithm \cite{lu2021data} is applied to crop the tissue from the background. Then the cropped images are split into patches and formulate as a WSI bag:
\begin{equation} \label{eq:wsi_bag}
X^p_i=\{x^{p}_{i,n}\}^{N_{p,i}}_{n=1},
\end{equation}
where $x^{p}_{i,n}$ represents a WSI patch and ${N_{p,i}}$ denotes the number of patches belongs to the $i$th patient.

\textbf{Genomics bags formulation.} Given that genomic data encompasses various attributes such as gene mutation status, RNA-Seq abundance and copy number variation, we follow \cite{liberzon2015molecular} and categorize the genes into six classes based on their functional roles: (i) tumor suppressor genes, (ii) oncogenes, (iii) protein kinases, (iv) cell differentiation markers, (v) transcription factors, and (vi) cytokines and growth factors. The genomic data in each class is concatenated into tensors, so that six bags of genomic data are obtained.
\begin{equation} \label{eq:genomic_bag}
X^g_i=\{x^{g}_{i,n}\}^{N_g}_{n=1},
\end{equation}
where $X^g_i$ denotes the whole bag of genomic data and $N_g$ represents the number of bags of genomic data, here $N_g=6$.

\textbf{Demographic and treatment bags formulation.} Since the demographics of the patient and the treatment information are discrete data, directly encoding them into the network will easily lead to optimization failures or sub-optimal performance. Inspired by large models like CLIP \cite{dosovitskiy2020image}, LLaMA \cite{genai2023llama} and BLIP \cite{li2022blip}, we embed discrete information by designing text prompts. For demographics, including sex, age and race of the patients, a simple text template is designed: ``{He/She} is a {age}-year-old {race} {man/woman}.". As treatment information represents whether radiation or pharmaceutical therapy was applied to the patients. Therefore, ``{Treatments} is/are applied." is adopted as the treatment template. Moreover, to enhance the feature representation of these two types of information, we also employ tensorized data to represent. So that, the demographic and treatment bags can be represented as
\begin{equation} \label{eq:text_bag}
X^t_i=\{x^{demo}_{text,i}, x^{treat}_{text,i}, x^{t}_{tensor,i}\},
\end{equation}
where $x^{demo}_{text,i}$ and $x^{treat}_{text,i}$ denote the generated demographic and treatment text for the $i$th patient, respectively, while $x^{t}_{tensor,i}\in \mathbb{R}^{5}$ is for the concatenated tensor array of demographic and treatment.

\subsection{Network}

\begin{figure*}[!t]
	\centering
	\includegraphics[scale=0.4]{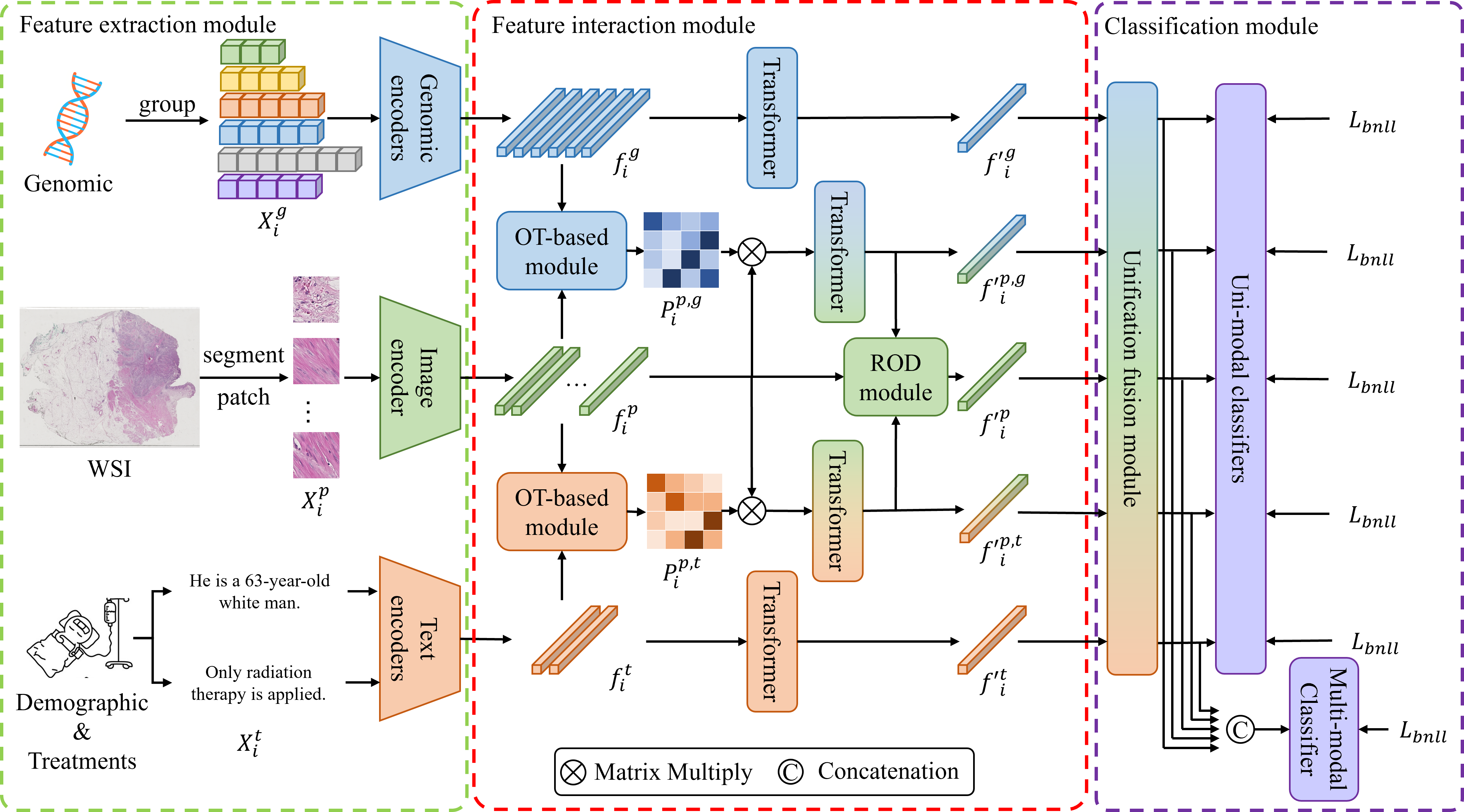}
	\caption{\label{fig:framework} The ICFNet consists of three different feature extractors, a feature interaction module and a classification module. To extract features from data in different modalities, an image encoder is utilized for WSI patches, a genomic encoder is utilized for genomic data, and CLIP-based text encoder \cite{gao2024clip} and MLP are adopted to encode demographic and treatment information.}
\end{figure*}

To extract and fuse above four types of information, we propose ICFNet, shown in Fig.\ref{fig:framework}, which consists of three components: a feature extractor, a multi-modal feature interaction and a classification module. The training and inference share the same algorithm flow.

\textbf{Feature extractors.} ICFNet incorporates multiple feature extractors. For histopathology WSIs, which are two-dimensional matrices with spatial relationships, a pretrained 2D convolution network is adopted to extract WSI features,
\begin{equation} \label{eq:wsi_feat}
f^{p}_{i}=F_{p}\left(x^{p}_{i,n}\right), n\in[1,N_{p,i}]
\end{equation}
where $F_{p}\left(\cdot\right)$ denotes the ImageNet-pretrained ResNet50\cite{he2016deep} and $f^{p}_{i} \in \mathbb{R}^{N_{p,i} \times C}$ is for the WSI feature of each patient, $N_{p,i}$ and $C$ represent the the number for patches in a bag and the extracted feature channels, respectively.

As for the tensorized genomic data, an off-the-shelf genomic encoder is leveraged to extract the genomic features. Given that each group of genes varies in quantity and function, we employ six separate self-normalizing neural networks (SNNs) \cite{klambauer2017self} with non-shared parameters to encode these six groups of genomic data independently:
\begin{equation} \label{eq:genomic_feat}
f^{g}_{i}=F_{g,n}\left(x^{g}_{i,n}\right), n\in[1,N_g]
\end{equation}
where $F_{g,n}\left(\cdot\right)$ denotes the trainable SNN for each group of genomic data, $f^{g}_{i} \in \mathbb{R}^{N_g \times C}$ represents the extracted genomic feature of each patient, $N_g=6$ in the experiments.

Meanwhile, a CLIP-based text encoder is adopted for demographic and treatment text feature extraction. And for the tensorized demographic and treatment data, multilayer perceptron (MLP) is applied. After that, the three features are stacked as 
\begin{equation} \label{eq:feat_bag}
f^{t}_{i}=stack\left(F_{clip}\left(x^{demo}_{text,i}\right), F_{clip}\left(x^{treat}_{text,i} \right), F_{mlp}\left(x^{t}_{tensor,i}\right) \right),
\end{equation}
where $F_{clip}\left(\cdot\right)$ and $F_{mlp}\left(\cdot\right)$ stand for CLIP-based text encoder and MLP respectively. $f^{t}_{i} \in \mathbb{R}^{N_t \times C}$ represents the extracted feature of demographic and treatment information and $N_t=3$ in the experiments.

\textbf{Multi-modal feature interaction module.} To fuse information across different modalities, optimal transport (OT) algorithm \cite{xu2023multimodal} is utilized to learn the optimal matching flow between the different features. As the histopathology patches contain the largest amount of information, only two modality interactions are considered, including the interaction between image feature and genomic feature, as well as the interaction between image feature and text feature. Specifically, a discrete Kantorovich formulation \cite{kantorovich2006translocation} is adopted to search the optimal general matching flow between the features:
\begin{equation} \label{eq:ot_matching_flow}
W\left(f^{p}_{i}, f^{X}_{i}\right)= \min \limits_{P^{p,X}_{i}\in \Pi(\mu_p,\mu_X)}<P^{p,X}_i,C^{p,X}_i>_F,
\end{equation}
where $W(\cdot)$ denotes the OT algorithm function, $P^{p,X}_i$ represents the optimal matching flow, $C^{p,X}_i \in \mathbb{R}^{M_p\times M_X}$ is the cost matrix provided by $C^{u,v}_i=dist(f^{p}_{i,u},f^{X}_{i,v})$ with a ground distance metric function $dist(\cdot)$, $\Pi(\mu_p,\mu_X)$ is to constrain the total mass equality between marginal distributions, and $X$ can be $g$ or $t$ for genomic or text features respectively. $<\cdot,\cdot>_F$ refers to the Frobenius dot product, so that Eq.\ref{eq:ot_matching_flow} is able to achieve the minimum cost matching flow based on pairwise similarity. After obtaining the optimal matching flow $P_n$ with OT-based module, the features from different modalities can be aligned with $f^{p,X}_i={P^{p,X}_i}^\top f^{p}_i$. The application of OT-based attention mechanisms not only extracts relevant information from image features to other modalities, but also keep the global structure consistency with other features, as well as accomplishes dimensional reduction of image features.

\begin{figure}[!t]
	\centering
	\includegraphics[scale=0.6]{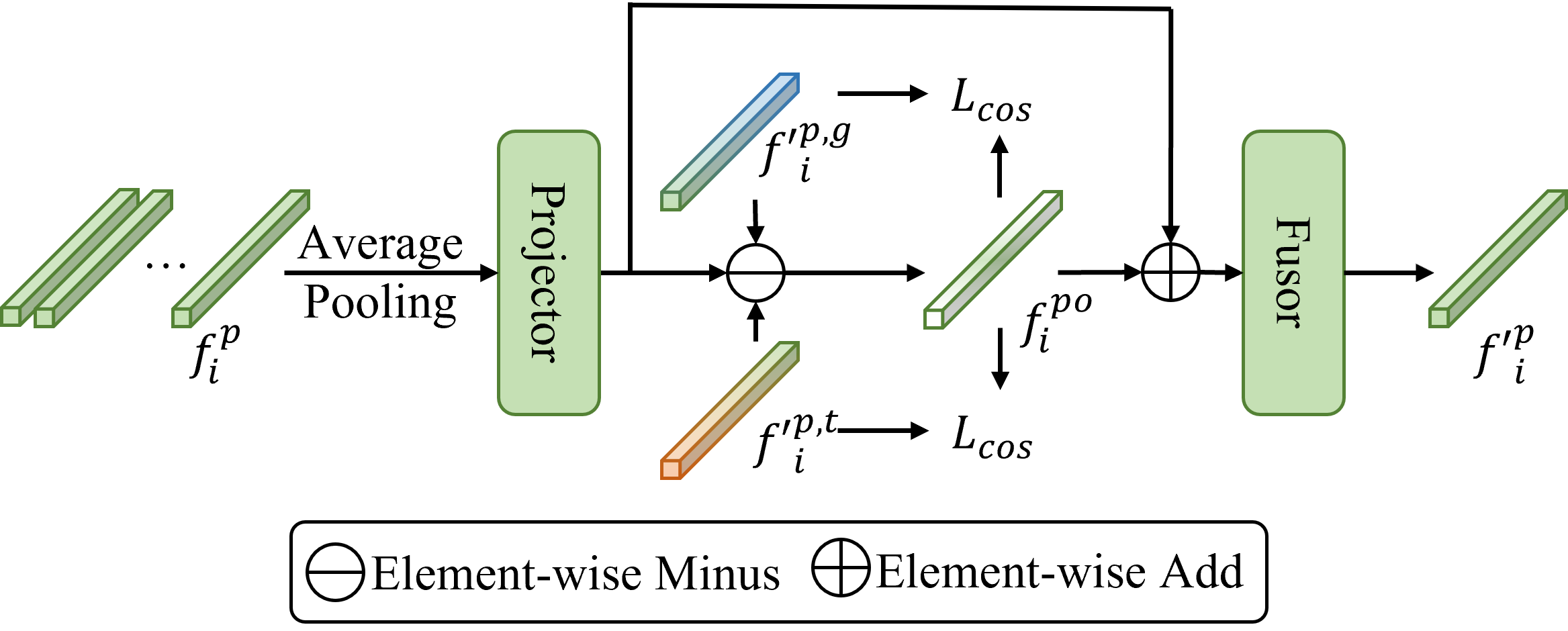}
	\caption{\label{fig:rod} The structure of ROD module. An average pooling and projector are applied to fuse the patch features, then the patch features are abstracted by histo-genomic and histo-text features to reduce information redundency. The cosine similarity loss is adopted to ensure that features are orthogonal to each other. To enhance the modality-specific feature, a residual structure and a fusor are adopted to fuse the features.}
\end{figure}

Acquiring all the features mentioned above, four Transformer-based modules are adopted to further extract and fuse intra-modality information for each feature, obtaining $f'^{g}_{i}$, $f'^{p,g}_{i}$, $f'^{p,t}_{i}$ and $f'^{t}_{i}$. To reduce the redundancy of information across modalities and enhance the modality-specific features, ROD module is proposed. As shown in Fig.\ref{fig:rod}, the patch features are firstly fused with an average pooling and a projector, then abstracted by the histo-genomic and histo-text features to reduce the redundency with modality-specific feature $f^{po}_{i}$ obtained. To maintain the patch information and modality-specific feature, a residual structure and a fusor is applied to fuse the features, obtaining the final image feature $f'^{p}_{i}$. The projector and the fusor are both consist of a linear layer with an activation function.

Finally, five different features are obtained for the classification.

\begin{figure}[!t]
	\centering
	\includegraphics[scale=0.6]{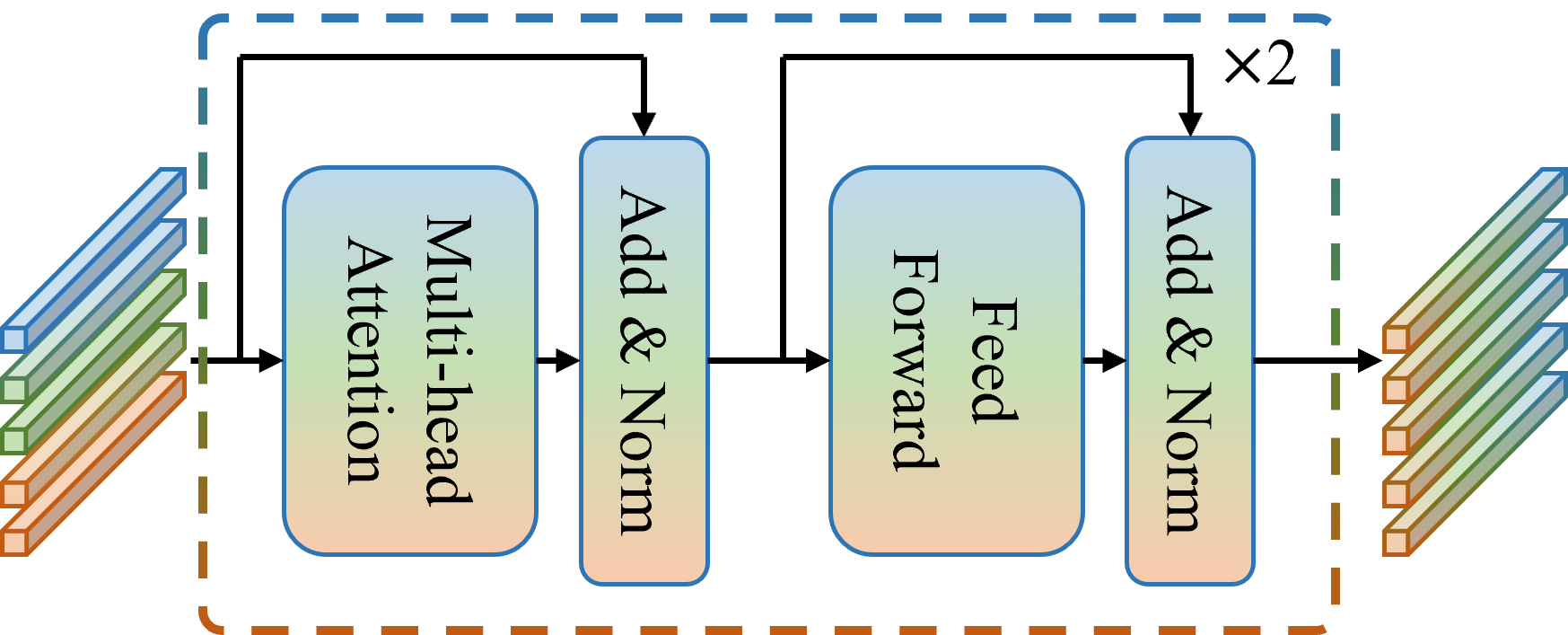}
	\caption{\label{fig:unimodule} The unification fusion module consists two multi-head attention based blocks. The module can map the five features into a unified latent space for further classification tasks.}
\end{figure}

\textbf{Classification module.} As the five distinct features originate from different modalities, they are likely to exist in disparate latent spaces. To address this issue, we employ a multi-head attention module to fuse them, as illustrated in Fig.\ref{fig:unimodule}, thereby mapping the various modalities into a unified latent space. Then, the five features are concatenated and fed into the multi-modal classifier. The classifier comprises a single linear layer to predict the independent probability of hazard for each patient in each survival time bin. 
\begin{equation} \label{eq:prediction}
S_{haz,i} = F_{cls}\left(concat\left(f''^g_i, f''^{p,g}_i, f''^p_i, f''^{p,t}_i, f''^t_i\right)\right),
\end{equation}
where $F_{cls}$ represents the multi-modal classifier, the $f''^X_i$ denotes the unified feature of each corresponding modality, and $S_{haz,i} \in \mathbf{R}^{N_b}_{+}$ is the predicted hazard scores of patients in survival time bins, while $N_b$ is the number of survival time bins predetermined based on the survival times of patients in the training set.

To enhance the network's learning ability of discriminative information from each feature, dense loss function scheme is employed. Specifically, for each feature, an independent classifier is utilized to predict patient survival, with supervision provided by the corresponding labels.

\subsection{Loss functions}
Due to the presence of censored data in prognostic datasets, NLLLoss \cite{zadeh2020bias} is widely adopted in prognostic tasks,
\begin{equation} \label{eq:nllloss}
\begin{aligned}
L_{nll} = & -c_i \cdot \mathrm{log}(S_{surv,i}(y_{i})) \\
          & -(1-c_i) \cdot \mathrm{log}(S_{surv,i}(y_{i}-1)) \\
          & -(1-c_i) \cdot \mathrm{log}(S_{haz,i}(y_{i})) \\
\end{aligned}
\end{equation}
in which $c_i$ represents the censorship of the patient, $S_{surv,i}(y_i)$ denotes the survival probability of the patient till the ground truth survival time bin $y_i$, and $S_{haz,i}(y_{i})$ stands for the hazard score of the patient in the $y_i$ time bin. As for $S_{surv,i}(y_i)$, it is a cumulative probability, which can be expressed as
\begin{equation} \label{eq:survival_score}
S_{surv,i}(y_i) = \prod_{n=0}^{y_i} (1-S_{haz,i,n}),
\end{equation}
in which $S_{haz,i,n}$ is the predicted hazard score for the $i$th patient in $n$th time bin.

However, because $S_{haz,i}$ is a value between 0 and 1, $S_{surv,i}$ tends to be disproportionately larger for higher values of $y_{i}$. This inherently causes the network to focus more on instances with larger $y_{i}$ (longer survival times) while inadequately training on patients with smaller $y_{i}$ (shorter survival times). To address this issue, we propose Balanced Negative Log-Likelihood Loss (BNLLLoss) to balance the weights between different classes by incorporating the loss of $S_{haz,i}$ into terms of $S_{surv,i}$ in NLLLoss,
\begin{equation} \label{eq:balanced_nllloss}
\begin{aligned}
&L_{bnll} = -c_i \cdot \mathrm{log}\left(S_{surv,i}(y_{i}) \cdot (1-S_{haz,i}(y_i))^{N_b-1-y_i} \right) \\
&-(1-c_i) \cdot \mathrm{log}\left(S_{surv,i}(y_{i}-1) \cdot (1-S_{haz,i}(y_i-1))^{N_b-1-y_i}\right) \\
&-(1-c_i) \cdot \mathrm{log}\left(S_{haz,i}(y_{i})\right), \\
\end{aligned}
\end{equation}
where $L_{bnll}$ represents the BNLLLoss function. Via BNLLLoss, each class can be trained with same weights. As a uni-modal classifier is utilized for each feature, BNLLLoss is also applied to supervise the uni-modal classifiers. 

Moreover, to ensure the redundancy can be reduced, a cosine similarity loss is adopted in the ROD module. Hence, the total loss function can be represented as
\begin{equation} \label{eq:total_loss}
L_{total} = L_{m} + \alpha \cdot (L_u^g+L_u^{p,g}+L_u^p+L_u^{p,t}+L_u^t+L_{cos}^{p,g} + L_{cos}^{p,t}), 
\end{equation}
where $L_m$ represents the loss for the multi-modal classifier, each $L_u^X$ denotes the loss for the corresponding uni-modal classifier, $L_{cos}^{p,g}$ and $L_{cos}^{p,t}$ represent the cosine similarity between $f'^{p,g}_{i}$ and $f^{po}_{i}$ as well as $f'^{p,t}_{i}$ and $f^{po}_{i}$, respectively, $\alpha$ and $\beta$ are hyper-parameters which are set as 0.1 to balance the model training in our experiments.

Finally, as the training and inference phase share the same procedure, we can obtain the risk scores by adding up all the $S_{surv,i}$
\begin{equation} \label{eq:risk}
S_{r,i} = -\sum_{n=1}^{N_b} S_{surv,i}(n),
\end{equation}
where $S_{r,i}$ is the predicted risk score for the $i$th patient. Patients with longer survival times are associated with lower risk, and vice versa.

\section{Experiments}
\label{sec:experiments}

\subsection{Datasets and evaluation metrics}
To evaluate the performance of ICFNet, a series of experiments are conducted over five open-source cancer datasets in TCGA, including Bladder Urothelial Carcinoma (BLCA), Breast Invasive Carcinoma (BRCA), Glioblastoma \& Lower Grade Glioma (GBMLGG), Lung Adenocarcinoma (LUAD), and Uterine Corpus Endometrial Carcinoma (UCEC). The details of the datasets are shown in Table \ref{table:datasets}. Due to the limited number of cases in these five datasets, cross-validation process is applied to evaluate our model. In addition, the concordance index (C-index) is selected as the metric. The C-index metric pairs patients and calculates the proportion of all pairs that are consistent with the relationship between the actual and predicted risk.
 
\begin{table*}[!t]
\caption{\label{table:datasets}The details of the datasets. Geno. refers to genomics, while TSG, ONC, PK, CDM, TF and CGF represent Tumor Suppressor Genes, Oncogenes, Protein Kinases, Cell Differentiation Markers, Transcription Factors and Cytokines \& Growth Factors respectively.}
\begin{center}
\begin{tabular}{cc|ccccc} 
\hline
\multicolumn{2}{c|}{Datasets}   & BLCA & BRCA & GBMLGG & LUAD & UCEC\\
\hline
\multicolumn{2}{c|}{Cases}      & 373  & 956  & 569    & 453  & 480   \\
\multicolumn{2}{c|}{Slides}     & 437  & 1017 & 1014   & 516  & 539   \\
\hline
\multirow{6}{*}{Geno.} & TSG    & 94   & 91   & 84     & 89   & 3     \\
                       & ONC    & 334  & 353  & 314    & 334  & 24    \\
                       & PK     & 521  & 553  & 498    & 534  & 21    \\
                       & CDM    & 468  & 480  & 415    & 471  & 22    \\
                       & TF     & 1496 & 1566 & 1396   & 1510 & 65    \\
                       & CGF    & 479  & 480  & 428    & 482  & 15    \\
\hline
\end{tabular}
\end{center}
\end{table*}

\subsection{Implementation details}
To ensure a fair comparison, we follow MOTCat \cite{xu2023multimodal} and conduct all experiments. To formulate the WSI patch bags, OTSU method is firstly applied to segment tissue regions, and then non-overlapping patches with the size of $256 \times 256$ are extracted from the tissue region over $20\times$ magnification. As for the image encoder, an ImageNet-pretrained ResNet50 is adopted to extract the WSI patch features before the training stage. Note that, to prevent the pre-trained model from extracting excessively high-dimensional features, we exclusively select the output from layer3 as the patch features. SNN \cite{klambauer2017self} is utilized as the genomic encoder to encode the genomic bag. CLIP-Adapter \cite{gao2024clip} is adopted as the text encoder and a two-layer MLP is adopted to encoder tensorized data. The CLIP-Adapter consists of a frozen CLIP model and an adapter for fine-tuning. In the training section, Adam optimizer is adopted, with the initial learning rate of $1\times 10^{-4}$ and the weight decay of $1\times 10^{-5}$. Due to the different number of patches in each WSI, the batch size is set as 1, while the gradient update step is set as 32 to accumulate the gradient. All the experiments are conducted on a server with the Ubuntu 18.04 LTS OS, an Intel CPU E5 2.20GHz CPU and one NVIDIA Tesla V100 GPU for training and testing. Python version is 3.9.19, CUDA version is 10.2 and PyTorch version is 1.12.1.

\subsection{Experimental results}

\textbf{Comparison with the state-of-the art methods.} Following the dataset partition in MOTCat \cite{xu2023multimodal}, we conduct the 5-fold experiments on the five datasets. To have fair comparsion with existing approaches, we reproduce some of them with the same setting and compare with the performance of ICFNet. As shown in Table \ref{table:results}, ICFNet achieves new state-of-the-art scores among the survival prediction approaches. On the five datasets, our ICFNet outperforms MOTCat \cite{xu2023multimodal} by 5.29\% C-index score on average. Specifically, ICFNet gains the improvement of 6.48\% on BLCA, 11.40\% on BRCA, 1.43\% on GBMLGG, 4.04\% on LUAD, and 3.28\% on UCEC.

By comparing the performance of models with different numbers of input modalities, it can be observed that the more modalities are included, the better performance the model can achieve. To fairly compare the performance of ICFNet with existing methods, demographic and treatment information are incorporated into MOTCat \cite{xu2023multimodal}, denoted as MOTCat+Text. Specifically, the text \& MLP encoders are introduced to encode the relevant information from demographic and treatment data, obtaining text features. However, these text features are not interacted with any other modality features. Instead, they are directly concatenated with other features before the classifier and used for classification. The performance of MOTCat+Text shows the effectiveness of the demographic and treatment for prognosis. However, due to the utilization of different encoders for each modality and the lack of interaction between the extracted features, MOTCat+Text can only obtain sub-optimal performance.

\begin{figure*}[!th]
	\centering
	\includegraphics[scale=0.43]{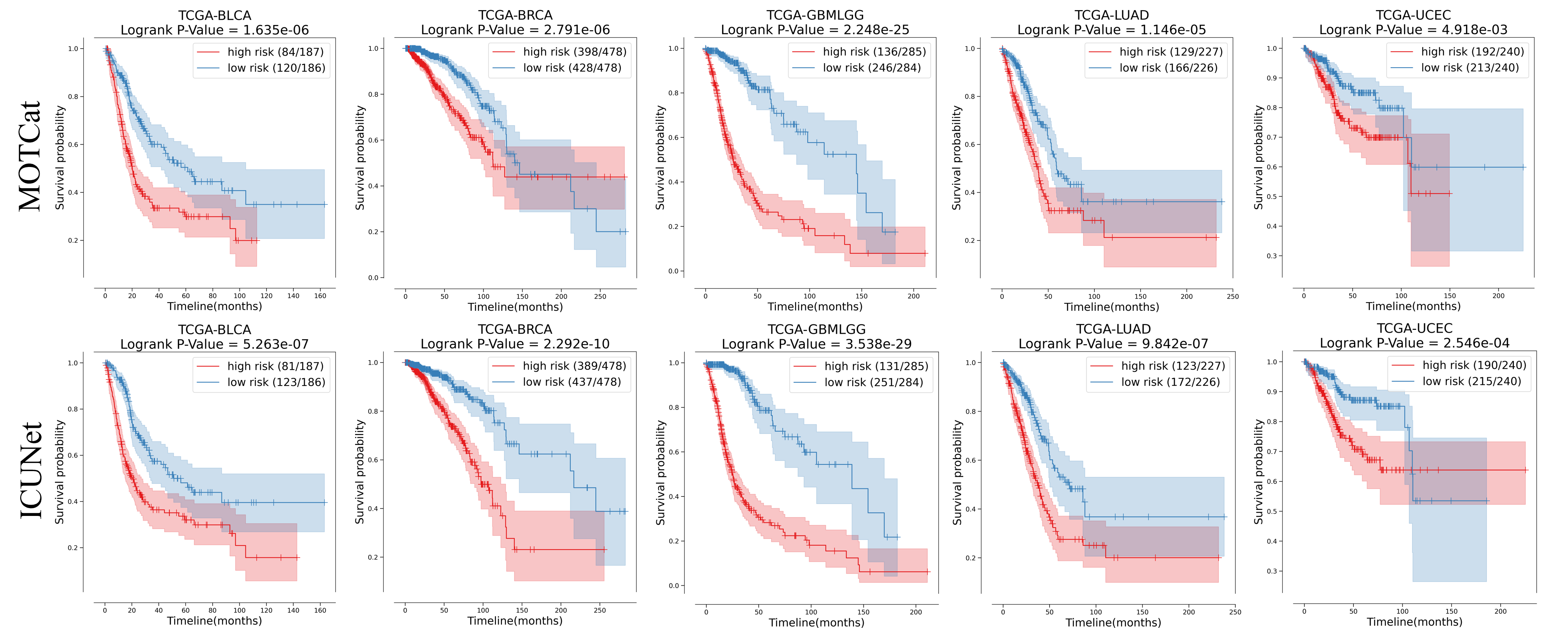}
	\caption{\label{fig:km_curve} The Kaplan-Meier survival curves of MOTCat \cite{xu2023multimodal} and ICFNet on five cancer datasets. The red and blue lines represent the high-risk and low-risk group, respectively. ICFNet exhibits larger margin between the two lines and smaller P-values than MOTCat.}
\end{figure*}

To better demonstrate the statistical difference of the ICFNet predictions for patients, the Kaplan-Meier (KM) survival curves are visualized for different approaches. Specifically, patients are divided into two groups, named as low-risk and high-risk, based on the predicted risk scores. Then, as shown in Fig.\ref{fig:km_curve} the statistics of ground truth survival time are presented for each group. Moreover, the Logrank test is conducted to measure the statistically significant difference between the two groups of patients. The lower P-value is obtained, the better performance of the model is indicated. Observing the Fig. \ref{fig:km_curve}, ICFNet gets a larger margin of between the low-risk and high-risk lines and achieves a lower P-value on all the datasets.

\begin{table*}[!t]
\caption{\label{table:results} The comparison of ICFNet with existing approaches on five TCGA datasets. ``P.", ``G." and ``T." denote the hispathological, genomic and text data, respectively. * represents that the metric is obtained via reproducing the related work. MOTCat+Text represents that the text \& MLP encoder is added to the MOTCat\cite{xu2023multimodal} but without any inter-modal fusion, then the text feature straightly concatenated with other feature before the classifier. The best and the second best results are highlighted in \textbf{bold} and in \underline{underline}, respectively.}
\centering
\resizebox{\linewidth}{!}{
\begin{tabular}{cccccccccc}
\hline
\multirow{2}{*}{Methods} & \multirow{2}{*}{P.} & \multirow{2}{*}{G.} & \multirow{2}{*}{T.} &\multicolumn{5}{c}{Datasets} & \multirow{2}{*}{Overall}\\
\cline{5-9}
 & & & & BLCA & BRCA & GBMLGG & LUAD & UCEC & \\
\hline
SNN \cite{klambauer2017self}                        &            & \checkmark & 
& 0.618$\pm$0.022 & 0.624$\pm$0.060 & 0.834$\pm$0.012 & 0.611$\pm$0.047 & 0.679$\pm$0.040 & 0.673 \\
SNNTrans \cite{klambauer2017self, shao2021transmil} &            & \checkmark & 
& 0.659$\pm$0.032 & 0.647$\pm$0.063 & 0.839$\pm$0.014 & 0.638$\pm$0.022 & 0.656$\pm$0.038 & 0.688 \\
\hline
AttnMIL \cite{ilse2018attention}                    & \checkmark &            &
& 0.551$\pm$0.049 & 0.577$\pm$0.043 & 0.786$\pm$0.026 & 0.561$\pm$0.078 & 0.639$\pm$0.057 & 0.623\\
DeepAttenMISL \cite{yao2020whole}                   & \checkmark &            &
& 0.504$\pm$0.042 & 0.524$\pm$0.043 & 0.734$\pm$0.029 & 0.548$\pm$0.050 & 0.597$\pm$0.059 & 0.581 \\
CLAM-SB \cite{lu2021data}                           & \checkmark &            &
& 0.559$\pm$0.034 & 0.573$\pm$0.044 & 0.779$\pm$0.031 & 0.594$\pm$0.063 & 0.644$\pm$0.061 & 0.630 \\
CLAM-MB \cite{lu2021data}                           & \checkmark &            &
& 0.565$\pm$0.027 & 0.578$\pm$0.032 & 0.776$\pm$0.034 & 0.582$\pm$0.072 & 0.609$\pm$0.082 & 0.622 \\
TransMIL \cite{shao2021transmil}                    & \checkmark &            &
& 0.608$\pm$0.139 & 0.626$\pm$0.042 & 0.798$\pm$0.033 & 0.641$\pm$0.033 & 0.657$\pm$0.044 & 0.666 \\
DTFD-MIL \cite{zhang2022dtfd}                       & \checkmark &            &
& 0.546$\pm$0.021 & 0.609$\pm$0.059 & 0.792$\pm$0.023 & 0.585$\pm$0.066 & 0.656$\pm$0.045 & 0.638 \\
Surformer \cite{wang2023surformer}                  & \checkmark &            &
& 0.594$\pm$0.027 & 0.628$\pm$0.037 & 0.809$\pm$0.026 & 0.591$\pm$0.064 & 0.681$\pm$0.028 & 0.661 \\
G-HANet \cite{wang2024histo}                        & \checkmark &            &
& 0.630$\pm$0.032 & 0.664$\pm$0.065 & 0.817$\pm$0.022 & 0.612$\pm$0.028 & \textbf{0.729$\pm$0.050} & 0.690 \\
\hline
Porpoise \cite{chen2022pan}                         & \checkmark & \checkmark &
& 0.636$\pm$0.024 & 0.652$\pm$0.042 & 0.834$\pm$0.017 & 0.647$\pm$0.031 & 0.695$\pm$0.032 & 0.693 \\
MCAT \cite{chen2021multimodal}                      & \checkmark & \checkmark &
& 0.672$\pm$0.032 & 0.659$\pm$0.031 & 0.835$\pm$0.024 & 0.659$\pm$0.027 & 0.649$\pm$0.043 & 0.695 \\
CMTA \cite{zhou2023cross}                           & \checkmark & \checkmark &
& \underline{0.691$\pm$0.043} & 0.668$\pm$0.043 & \underline{0.853$\pm$0.012} & \underline{0.686$\pm$0.036} & 0.698$\pm$0.041 & \underline{0.719} \\
MOTCat* \cite{xu2023multimodal}                     & \checkmark & \checkmark &
& 0.664$\pm$0.031 & 0.658$\pm$0.012 & 0.839$\pm$0.027 & 0.668$\pm$0.036 & 0.670$\pm$0.038 & 0.699 \\
SurvPath* \cite{jaume2024modeling}                  & \checkmark & \checkmark &
& 0.614$\pm$0.061 & 0.656$\pm$0.036 & 0.795$\pm$0.025 & 0.620$\pm$0.036 & 0.691$\pm$0.024 & 0.675 \\
CCL*\cite{zhou2024cohort}                            & \checkmark & \checkmark &
& 0.649$\pm$0.037 & 0.643$\pm$0.035 & 0.842$\pm$0.025 & 0.669$\pm$0.032 & 0.671$\pm$0.048 &  0.695 \\
MMP*\cite{Song2024Multimodal}                       & \checkmark & \checkmark &
& 0.640$\pm$0.029 & 0.681$\pm$0.042 & 0.850$\pm$0.020 & 0.614$\pm$0.051 & 0.572$\pm$0.050 & 0.671  \\
\hline
MOTCat+Text \cite{xu2023multimodal, dosovitskiy2020image} & \checkmark & \checkmark & \checkmark
& 0.667$\pm$0.033 & \underline{0.713$\pm$0.036} & 0.834$\pm$0.024 & 0.676$\pm$0.048 & 0.667$\pm$0.016 & 0.711 \\
ICFNet(Ours)                                        & \checkmark & \checkmark & \checkmark
& \textbf{0.709$\pm$0.044} & \textbf{0.724$\pm$0.041} & \textbf{0.854$\pm$0.023} & \textbf{0.697$\pm$0.044} & \underline{0.702$\pm$0.030} & \textbf{0.737} \\
\hline
\end{tabular}}
\end{table*}

\textbf{Ablation study of ICFNet.} To further explore the effectiveness of different ICFNet parts, we conduct a series of ablation experiments and the results are shown in Table \ref{table:ablation_module}. Via adopting OT-based attention mechanism instead of Transformer, it demonstrates a more robust capability for feature projection. The dense supervision and BNLL loss function facilitate better convergence of the network. By employing unification fusion module and ROD module, feature fusion can be enhanced while effectively reducing redundant information between features. These experimental results demonstrate the effectiveness of our proposed methods, which ensure the comprehensive utilization of demographic and treatment information while ensuring the sufficient utilization of other modalities as well.

\begin{table*}[!t]
\caption{\label{table:ablation_module} The ablation study of ICFNet with different modules. ``OT." stands for adopting the OT-based attention to fuse the features, instead of Transformer. ``Den." represents adopting dense loss for the uni-modal classifiers. ``BNLL." denotes whether the balanced nllloss replaces the nllloss. ``U.M." is for adopting the unification fusion module to unify the features. ``ROD" means adopting ROD module to reduce the feature redundency. The best results are highlighted in \textbf{bold}.}
\centering
\resizebox{\linewidth}{!}{
\begin{tabular}{ccccccccccc}
\hline
\multirow{2}{*}{OT.} & \multirow{2}{*}{Den.} & \multirow{2}{*}{BNLL.} & \multirow{2}{*}{U.M.} & \multirow{2}{*}{ROD.}&\multicolumn{5}{c}{Datasets} & \multirow{2}{*}{Overall}\\
\cline{6-10}
 & & & & & BLCA & BRCA & GBMLGG & LUAD & UCEC & \\
\hline
           &            &            &            &
& 0.682$\pm$0.023 & 0.704$\pm$0.027 & 0.843$\pm$0.021 & 0.678$\pm$0.023 & 0.673$\pm$0.018 & 0.716 \\           
\checkmark &            &            &            &
& 0.680$\pm$0.021 & 0.718$\pm$0.027 & 0.845$\pm$0.026 & 0.674$\pm$0.027 & 0.667$\pm$0.024 & 0.717 \\
\checkmark & \checkmark &            &            &
& 0.683$\pm$0.021 & 0.722$\pm$0.027 & 0.846$\pm$0.023 & 0.677$\pm$0.057 & 0.675$\pm$0.042 & 0.721 \\
\checkmark & \checkmark & \checkmark &            &
& 0.690$\pm$0.023 & 0.710$\pm$0.038 & 0.844$\pm$0.017 & 0.687$\pm$0.052 & 0.681$\pm$0.039 & 0.722 \\
\checkmark & \checkmark &            & \checkmark & 
& 0.695$\pm$0.016 & 0.721$\pm$0.043 & 0.845$\pm$0.025 & 0.685$\pm$0.041 & 0.684$\pm$0.038 & 0.726 \\
\checkmark &            & \checkmark & \checkmark & 
& 0.707$\pm$0.016 & 0.729$\pm$0.024 & 0.843$\pm$0.025 & 0.678$\pm$0.045 & 0.685$\pm$0.023 & 0.728 \\
           & \checkmark & \checkmark & \checkmark & 
& 0.689$\pm$0.038 & 0.722$\pm$0.0.007 & 0.838$\pm$0.023 & \textbf{0.711$\pm$0.045} & 0.668$\pm$0.033 & 0.726 \\
\checkmark & \checkmark & \checkmark & \checkmark & 
& 0.707$\pm$0.016 & \textbf{0.729$\pm$0.024} & 0.843$\pm$0.017 & 0.690$\pm$0.054 & 0.685$\pm$0.023 & 0.731 \\
\checkmark & \checkmark & \checkmark & \checkmark & \checkmark
& \textbf{0.709$\pm$0.044} & 0.724$\pm$0.041 & \textbf{0.854$\pm$0.023} & 0.697$\pm$0.044 & \textbf{0.702$\pm$0.030} & \textbf{0.737} \\
\hline
\end{tabular}}
\end{table*}

\begin{table*}[!t]
\caption{\label{table:ablation_modal} The ablation study of ICFNet with different information. ``Demo" refers to demographic information, while ``Treat" represents to treatment information. ``Tensor" denotes if the tensorized array is adopted.}
\centering
\resizebox{\linewidth}{!}{
\begin{tabular}{cccccccccc}
\hline
\multirow{2}{*}{Modality} & \multirow{2}{*}{Demo} & \multirow{2}{*}{Treat} & \multirow{2}{*}{Tensor} &\multicolumn{5}{c}{Datasets} & \multirow{2}{*}{Overall}\\
\cline{5-9}
 & & & & BLCA & BRCA & GBMLGG & LUAD & UCEC & \\
\hline
\multirow{5}{*}{Multi-modal}
&            &            & 
& 0.663$\pm$0.030 & 0.663$\pm$0.013 & 0.832$\pm$0.022 & 0.668$\pm$0.057 & 0.665$\pm$0.051 & 0.698 \\
& \checkmark &            & 
& 0.679$\pm$0.021 & 0.670$\pm$0.023 & 0.839$\pm$0.024 & 0.681$\pm$0.033 & 0.659$\pm$0.048 & 0.706 \\
&            & \checkmark & 
& 0.695$\pm$0.018 & 0.714$\pm$0.037 & 0.849$\pm$0.035 & 0.690$\pm$0.047 & 0.679$\pm$0.033 & 0.725 \\
& \checkmark & \checkmark & 
& 0.693$\pm$0.025 & 0.714$\pm$0.042 & 0.839$\pm$0.027 & 0.697$\pm$0.045 & 0.682$\pm$0.031 & 0.725 \\
& \checkmark & \checkmark & \checkmark
& \textbf{0.709$\pm$0.044} & \textbf{0.724$\pm$0.041} & \textbf{0.854$\pm$0.023} & \textbf{0.697$\pm$0.044} & \textbf{0.702$\pm$0.030} & \textbf{0.737} \\
\hline
\multirow{4}{*}{Text-only}
& \checkmark &            & 
& 0.585$\pm$0.042 & 0.635$\pm$0.070 & 0.725$\pm$0.042 & 0.556$\pm$0.072 & 0.583$\pm$0.075 & 0.617 \\
&            & \checkmark & 
& 0.582$\pm$0.015 & 0.672$\pm$0.049 & 0.556$\pm$0.027 & 0.549$\pm$0.062 & 0.599$\pm$0.051 & 0.592 \\
& \checkmark & \checkmark & 
& 0.621$\pm$0.051 & 0.700$\pm$0.074 & 0.738$\pm$0.031 & 0.591$\pm$0.042 & 0.633$\pm$0.046 & 0.657 \\
& \checkmark & \checkmark & \checkmark
& 0.610$\pm$0.034 & 0.705$\pm$0.068 & 0.719$\pm$0.025 & 0.578$\pm$0.069 & 0.610$\pm$0.076 & 0.644 \\
\hline
\end{tabular}}
\end{table*}

Furthermore, to evaluate the contribution of demographic and treatment information, we conduct experiments by gradually removing these information. As illustrated in Table \ref{table:ablation_modal}, both of the demographic and treatment information have positive influence to the model. Due to the variation in age, gender and racial distribution among different cancer patient populations, as well as the varying effectiveness of treatment methods used, it is essential to quantify the effectiveness of demographic and treatment information for different types of cancer. Therefore, the text-only experiments are conducted, which take the demographic and treatment information as the only input, and remove the WSI and genomic features as well as the corresponding branches. The results indicate the upper bound of the demographic and treatment information in prognosis and exhibit the effectiveness of multi-modal fusion. For example, the text-only model achieves promising performance on BRCA, indicating that text information is effective in predicting the survival time of breast cancer patients. Conversely, its performance on LUAD is poorer, suggesting that accurate survival prediction for lung cancer patients is challenging based solely on textual information. It is evident that ICFNet selects features from other modalities that are more advantageous for survival prediction, while ensuring model performance. On the other hand, the tensorized array can enhance the model performance. It can be observed that the tensorized demographic and treatment information leads to a performance decline in the text-only model but improves the performance of the multi-modal models. We attribute this improvement to the comprehensive characterization of patients provided by the multi-modal data. By describing patients from multiple perspectives, we not only enhance the accuracy of the patient assessment but also strengthen the feature extraction capability of each modality, so that improves the robustness of the model.

\subsection{Visualization}

\begin{figure*}[!th]
	\centering
	\includegraphics[scale=0.4]{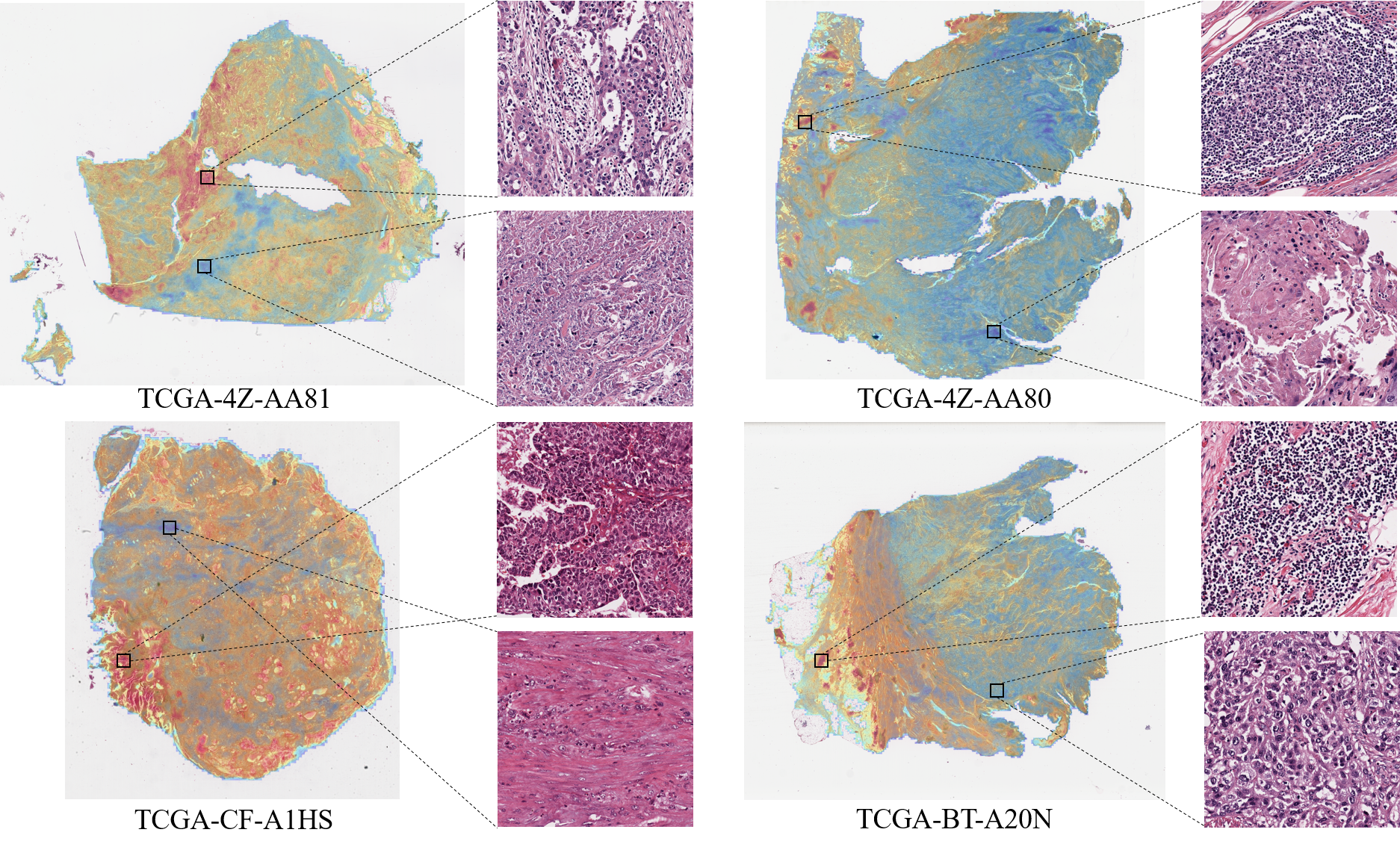}
	\caption{\label{fig:visualization} The visualization of CAM from ICFNet. The red region represents the area contributing the most to the patient's risk score, while the blue region indicates the opposite.}
\end{figure*}

To demonstrate the interpretability of the model, GradCAM \cite{selvaraju2017grad} is utilized to visualize the class activation maps (CAM) of certain slides. As shown in Fig.\ref{fig:visualization}, the red region represents the area contributing the most to the patient's risk score, while the blue region indicates the opposite. By observing the morphologies of tissue in these red regions, we can summarize the prototypes that lead to high patient risk, facilitating rapid analysis by experts.

\subsection{Discussion}
\textbf{The value of ICFNet for clinical decision-making.} Different from previous prognostic models that rely solely on histopathology or genomic expression features, ICFNet integrates patient demographics and treatment protocols to provide a more comprehensive evaluation, thereby enhancing prognostic performance. It's important to note that for clinicians, a patient's existing information, such as histopathology, genomics, and demographics, is fixed, whereas treatment protocols are variable. This means that during clinical decision-making, clinicians can adjust the text input for ICFNet to observe differences in patient prognosis, ultimately identifying treatment plans that yield better prognostic outcomes and provide a more quantifiable basis for treatment decisions.

\begin{figure*}[!t]
	\centering
	\includegraphics[scale=0.4]{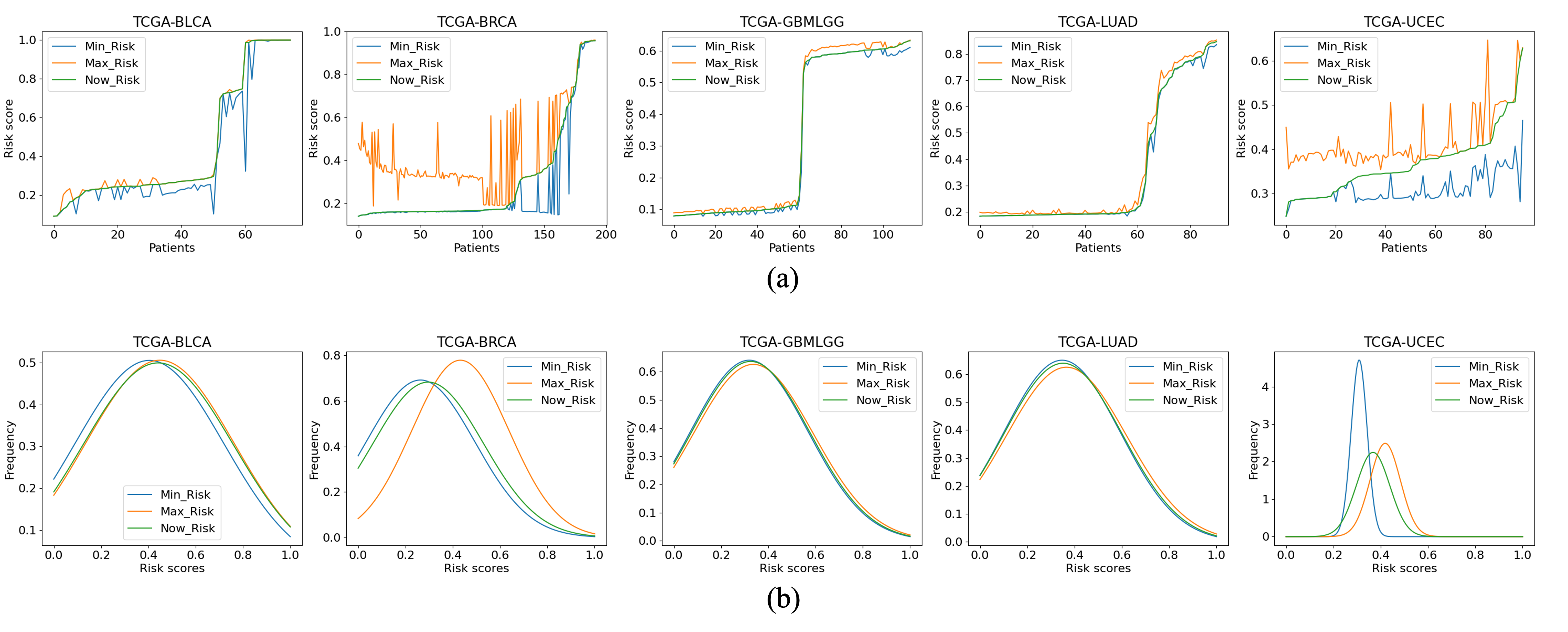}
	\caption{\label{fig:variation} The analysis of the survival predictions on the related datasets. (a) is the survival prediction for each patient under various treatment methods. (b) is the normalized probability density of the predicted risk scores.}
\end{figure*}

To illustrate the impact of different treatment protocols on prognosis, Fig.\ref{fig:variation} shows the range of prognostic variations for Fold 0 from the related datasets when different treatment options are applied. Fig.\ref{fig:variation}(a) presents the survival predictions for each patient under various treatment methods as given by ICFNet. Fig.\ref{fig:variation}(b) shows the probability density estimates obtained by fitting the survival prediction results to a normal distribution. In this figure, the green line represents the survival predictions for the actual treatment protocols used in the dataset, while the orange and blue lines represent the best and worst prognostic outcomes under different treatment protocols, respectively. Fig.\ref{fig:variation} demonstrates, both locally and globally, that treatment protocols significantly impact prognosis. This impact indicates that the treatment decisions made at the time, possibly constrained by patient preferences and medical conditions, might not be optimal for the best prognostic outcome. However, clinicians can use ICFNet to directly observe survival estimates, making clinical decisions more quantifiable and efficient.

\textbf{Reasons for discrepancy between ICFNet's survival predictions and actual survival times.} Plenty of factors affect the accuracy of survival predictions. Intrinsic factors, such as individual heterogeneity, can influence disease progression rates and the effectiveness of medical interventions. Additionally, patient-specific conditions like comorbidities and psychological health may impact the accuracy of prognosis for a single disease. Extrinsic factors, including the patient's living environment and adherence to treatment protocols, also play a substantial role in creating variability in survival estimates. In summary, survival prediction is inclined towards higher probability events when dealing with large sample sizes, aligning with the core principles of data-driven artificial intelligence algorithms. Although there is still some discrepancy compared to actual survival times, ICFNet, which accounts for a broader range of factors, has achieved state-of-the-art prognostic performance.

\textbf{Limitation and future works.} Although ICFNet has shown improvement compared to existing approaches, there is still significant room for further enhancements. Firstly, the processing of WSIs in ICFNet is relatively simplistic, while features extracted by ImageNet pre-trained ResNet50 may not necessarily be the most informative for survival prediction. Secondly, although we have incorporated patient demographic and treatment information, there are still numerous factors that can influence patient’s survival, such as medical conditions, physical status and psychological well-being. Therefore, in future work, we attempt to develop a learnable WSI feature extractor to capture the most relevant features for survival prediction, as well as explore the inclusion of additional patient information, such as diagnostic records and psychological assessments.

\section{Conclusion}
\label{sec:conclusion}
In this paper, a multi-modal prognosis framework is proposed, named as ICFNet, which achieves the state-of-the-art performance on multiple modality datasets. ICFNet incorporates four types of information, including histopathology WSIs, genomics expression, patient demographic and treatment profiles. Firstly, the MIL strategy is used to formulate the bags for training, while text templates and prompts are designed for demographic and treatment information. Secondly, three different encoders are adopted to extract features from different modalities. Then, the features are interacted with OT based modules to explore the relationship between the features. To reduce redundancy among modality and enhance modality-specific features, ROD module is proposed. Furthermore, we design a unification fusion module to align features and train the network using a dense loss function. Finally, BNLLLoss is introduced to ensure the network fairly treats with every patient. The experimental results indicate the effectiveness of our ICFNet, which can conduct a prior assessment of the treatment administered to patients, thereby avoiding over-treatment or wastage of resources.

\bibliographystyle{ieeetr}
\bibliography{refs.bib}

\begin{thebibliography}{10}

\bibitem{zhang2024mtcsnet}
B.~Zhang, Z.~Meng, H.~Li, Z.~Zhao, and F.~Su, ``Mtcsnet: One-stage learning and two-point labeling are sufficient for cell segmentation,'' {\em IEEE Transactions on Medical Imaging}, 2024.

\bibitem{meng2024nusea}
Z.~Meng, J.~Dong, B.~Zhang, S.~Li, R.~Wu, F.~Su, G.~Wang, L.~Guo, and Z.~Zhao, ``Nusea: Nuclei segmentation with ellipse annotations,'' {\em IEEE Journal of Biomedical and Health Informatics}, 2024.

\bibitem{lu2021data}
M.~Y. Lu, D.~F. Williamson, T.~Y. Chen, R.~J. Chen, M.~Barbieri, and F.~Mahmood, ``Data-efficient and weakly supervised computational pathology on whole-slide images,'' {\em Nature biomedical engineering}, vol.~5, no.~6, pp.~555--570, 2021.

\bibitem{vaswani2017attention}
A.~Vaswani, N.~Shazeer, N.~Parmar, J.~Uszkoreit, L.~Jones, A.~N. Gomez, {\L}.~Kaiser, and I.~Polosukhin, ``Attention is all you need,'' {\em Advances in neural information processing systems}, vol.~30, 2017.

\bibitem{dosovitskiy2020image}
A.~Dosovitskiy, L.~Beyer, A.~Kolesnikov, D.~Weissenborn, X.~Zhai, T.~Unterthiner, M.~Dehghani, M.~Minderer, G.~Heigold, S.~Gelly, {\em et~al.}, ``An image is worth 16x16 words: Transformers for image recognition at scale,'' in {\em International Conference on Learning Representations}, 2020.

\bibitem{klambauer2017self}
G.~Klambauer, T.~Unterthiner, A.~Mayr, and S.~Hochreiter, ``Self-normalizing neural networks,'' {\em Advances in neural information processing systems}, vol.~30, 2017.

\bibitem{lin2019ghrelin}
T.-C. Lin, Y.-M. Yeh, W.-L. Fan, Y.-C. Chang, W.-M. Lin, T.-Y. Yang, and M.~Hsiao, ``Ghrelin upregulates oncogenic aurora a to promote renal cell carcinoma invasion,'' {\em Cancers}, vol.~11, no.~3, p.~303, 2019.

\bibitem{wang2019mmp}
Q.-M. Wang, L.~Lv, Y.~Tang, L.~Zhang, and L.-F. Wang, ``Mmp-1 is overexpressed in triple-negative breast cancer tissues and the knockdown of mmp-1 expression inhibits tumor cell malignant behaviors in vitro,'' {\em Oncology letters}, vol.~17, no.~2, pp.~1732--1740, 2019.

\bibitem{chen2021multimodal}
R.~J. Chen, M.~Y. Lu, W.-H. Weng, T.~Y. Chen, D.~F. Williamson, T.~Manz, M.~Shady, and F.~Mahmood, ``Multimodal co-attention transformer for survival prediction in gigapixel whole slide images,'' in {\em Proceedings of the IEEE/CVF international conference on computer vision}, pp.~4015--4025, 2021.

\bibitem{xu2023multimodal}
Y.~Xu and H.~Chen, ``Multimodal optimal transport-based co-attention transformer with global structure consistency for survival prediction,'' in {\em Proceedings of the IEEE/CVF International Conference on Computer Vision}, pp.~21241--21251, 2023.

\bibitem{zadeh2020bias}
S.~G. Zadeh and M.~Schmid, ``Bias in cross-entropy-based training of deep survival networks,'' {\em IEEE transactions on pattern analysis and machine intelligence}, vol.~43, no.~9, pp.~3126--3137, 2020.

\bibitem{ilse2018attention}
M.~Ilse, J.~Tomczak, and M.~Welling, ``Attention-based deep multiple instance learning,'' in {\em International conference on machine learning}, pp.~2127--2136, PMLR, 2018.

\bibitem{raju2020graph}
A.~Raju, J.~Yao, M.~M. Haq, J.~Jonnagaddala, and J.~Huang, ``Graph attention multi-instance learning for accurate colorectal cancer staging,'' in {\em Medical Image Computing and Computer Assisted Intervention--MICCAI 2020: 23rd International Conference, Lima, Peru, October 4--8, 2020, Proceedings, Part V 23}, pp.~529--539, Springer, 2020.

\bibitem{sharma2021cluster}
Y.~Sharma, A.~Shrivastava, L.~Ehsan, C.~A. Moskaluk, S.~Syed, and D.~Brown, ``Cluster-to-conquer: A framework for end-to-end multi-instance learning for whole slide image classification,'' in {\em Medical Imaging with Deep Learning}, pp.~682--698, PMLR, 2021.

\bibitem{li2021dual}
B.~Li, Y.~Li, and K.~W. Eliceiri, ``Dual-stream multiple instance learning network for whole slide image classification with self-supervised contrastive learning,'' in {\em Proceedings of the IEEE/CVF conference on computer vision and pattern recognition}, pp.~14318--14328, 2021.

\bibitem{shao2021transmil}
Z.~Shao, H.~Bian, Y.~Chen, Y.~Wang, J.~Zhang, X.~Ji, {\em et~al.}, ``Transmil: Transformer based correlated multiple instance learning for whole slide image classification,'' {\em Advances in neural information processing systems}, vol.~34, pp.~2136--2147, 2021.

\bibitem{zhang2022dtfd}
H.~Zhang, Y.~Meng, Y.~Zhao, Y.~Qiao, X.~Yang, S.~E. Coupland, and Y.~Zheng, ``Dtfd-mil: Double-tier feature distillation multiple instance learning for histopathology whole slide image classification,'' in {\em Proceedings of the IEEE/CVF conference on computer vision and pattern recognition}, pp.~18802--18812, 2022.

\bibitem{yan2023sparse}
R.~Yan, Z.~Lv, Z.~Yang, S.~Lin, C.~Zheng, and F.~Zhang, ``Sparse and hierarchical transformer for survival analysis on whole slide images,'' {\em IEEE Journal of Biomedical and Health Informatics}, 2023.

\bibitem{chen2023dmil}
Y.~Chen, Z.~Shao, H.~Bian, Z.~Fang, Y.~Wang, Y.~Cai, H.~Wang, G.~Liu, X.~Li, and Y.~Zhang, ``dmil-transformer: Multiple instance learning via integrating morphological and spatial information for lymph node metastasis classification,'' {\em IEEE Journal of Biomedical and Health Informatics}, vol.~27, no.~9, pp.~4433--4443, 2023.

\bibitem{sun2023tgmil}
X.~Sun, W.~Li, B.~Fu, Y.~Peng, J.~He, L.~Wang, T.~Yang, X.~Meng, J.~Li, J.~Wang, {\em et~al.}, ``Tgmil: A hybrid multi-instance learning model based on the transformer and the graph attention network for whole-slide images classification of renal cell carcinoma,'' {\em Computer Methods and Programs in Biomedicine}, vol.~242, p.~107789, 2023.

\bibitem{deng2024cross}
R.~Deng, C.~Cui, L.~W. Remedios, S.~Bao, R.~M. Womick, S.~Chiron, J.~Li, J.~T. Roland, K.~S. Lau, Q.~Liu, {\em et~al.}, ``Cross-scale multi-instance learning for pathological image diagnosis,'' {\em Medical image analysis}, vol.~94, p.~103124, 2024.

\bibitem{tsiknakis2024unveiling}
N.~Tsiknakis, G.~Manikis, E.~Tzoras, D.~Salgkamis, J.~M. Vidal, K.~Wang, D.~Zaridis, E.~Sifakis, I.~Zerdes, J.~Bergh, {\em et~al.}, ``Unveiling the power of model-agnostic multiscale analysis for enhancing artificial intelligence models in breast cancer histopathology images,'' {\em IEEE Journal of Biomedical and Health Informatics}, 2024.

\bibitem{kapse2024si}
S.~Kapse, P.~Pati, S.~Das, J.~Zhang, C.~Chen, M.~Vakalopoulou, J.~Saltz, D.~Samaras, R.~R. Gupta, and P.~Prasanna, ``Si-mil: Taming deep mil for self-interpretability in gigapixel histopathology,'' in {\em Proceedings of the IEEE/CVF Conference on Computer Vision and Pattern Recognition}, pp.~11226--11237, 2024.

\bibitem{yao2020whole}
J.~Yao, X.~Zhu, J.~Jonnagaddala, N.~Hawkins, and J.~Huang, ``Whole slide images based cancer survival prediction using attention guided deep multiple instance learning networks,'' {\em Medical Image Analysis}, vol.~65, p.~101789, 2020.

\bibitem{chen2021whole}
R.~J. Chen, M.~Y. Lu, M.~Shaban, C.~Chen, T.~Y. Chen, D.~F. Williamson, and F.~Mahmood, ``Whole slide images are 2d point clouds: Context-aware survival prediction using patch-based graph convolutional networks,'' in {\em Medical Image Computing and Computer Assisted Intervention--MICCAI 2021: 24th International Conference, Strasbourg, France, September 27--October 1, 2021, Proceedings, Part VIII 24}, pp.~339--349, Springer, 2021.

\bibitem{sandarenu2022survival}
P.~Sandarenu, E.~K. Millar, Y.~Song, L.~Browne, J.~Beretov, J.~Lynch, P.~H. Graham, J.~Jonnagaddala, N.~Hawkins, J.~Huang, {\em et~al.}, ``Survival prediction in triple negative breast cancer using multiple instance learning of histopathological images,'' {\em Scientific Reports}, vol.~12, no.~1, p.~14527, 2022.

\bibitem{wang2023surformer}
Z.~Wang, Q.~Gao, X.~Yi, X.~Zhang, Y.~Zhang, D.~Zhang, P.~Li{\`o}, C.~Bain, R.~Bassed, S.~Li, {\em et~al.}, ``Surformer: An interpretable pattern-perceptive survival transformer for cancer survival prediction from histopathology whole slide images,'' {\em Computer Methods and Programs in Biomedicine}, vol.~241, p.~107733, 2023.

\bibitem{shao2023lnpl}
Z.~Shao, Y.~Wang, Y.~Chen, H.~Bian, S.~Liu, H.~Wang, and Y.~Zhang, ``Lnpl-mil: Learning from noisy pseudo labels for promoting multiple instance learning in whole slide image,'' in {\em Proceedings of the IEEE/CVF International Conference on Computer Vision}, pp.~21495--21505, 2023.

\bibitem{shao2024multi}
W.~Shao, H.~Shi, J.~Liu, Y.~Zuo, L.~Sun, T.~Xia, W.~Chen, P.~Wan, J.~Sheng, Q.~Zhu, {\em et~al.}, ``Multi-instance multi-task learning for joint clinical outcome and genomic profile predictions from the histopathological images,'' {\em IEEE Transactions on Medical Imaging}, 2024.

\bibitem{chen2020pathomic}
R.~J. Chen, M.~Y. Lu, J.~Wang, D.~F. Williamson, S.~J. Rodig, N.~I. Lindeman, and F.~Mahmood, ``Pathomic fusion: an integrated framework for fusing histopathology and genomic features for cancer diagnosis and prognosis,'' {\em IEEE Transactions on Medical Imaging}, vol.~41, no.~4, pp.~757--770, 2020.

\bibitem{he2020feasibility}
Q.~He, X.~Li, D.~N. Kim, X.~Jia, X.~Gu, X.~Zhen, and L.~Zhou, ``Feasibility study of a multi-criteria decision-making based hierarchical model for multi-modality feature and multi-classifier fusion: Applications in medical prognosis prediction,'' {\em Information Fusion}, vol.~55, pp.~207--219, 2020.

\bibitem{jaume2024modeling}
G.~Jaume, A.~Vaidya, R.~J. Chen, D.~F. Williamson, P.~P. Liang, and F.~Mahmood, ``Modeling dense multimodal interactions between biological pathways and histology for survival prediction,'' in {\em Proceedings of the IEEE/CVF Conference on Computer Vision and Pattern Recognition}, pp.~11579--11590, 2024.

\bibitem{zhou2023cross}
F.~Zhou and H.~Chen, ``Cross-modal translation and alignment for survival analysis,'' in {\em Proceedings of the IEEE/CVF International Conference on Computer Vision}, pp.~21485--21494, 2023.

\bibitem{Zhang2024PROTOTYPICAL}
Y.~Zhang, Y.~Xu, J.~Chen, F.~Xie, and H.~Chen, ``Prototypical information bottlenecking and disentangling for multimodal cancer survival prediction,'' (Hybrid, Vienna, Austria), 2024.
\newblock Discriminability;Genomic data;Image data;Information bottleneck;Multi-modal;Multi-modal data;Multi-modal learning;Pathological images;Survival prediction;Whole slide images;.

\bibitem{Song2024Multimodal}
A.~H. Song, R.~J. Chen, G.~Jaume, A.~Vaidya, A.~S. Baras, and F.~Mahmood, ``Multimodal prototyping for cancer survival prediction,'' vol.~235, (Vienna, Austria), pp.~46050 -- 46073, 2024.
\newblock 'current;Gene groups;Interpretability;Multi-modal;Multimodal prototyping;Small patches;Survival prediction;Tokenizing;Transcriptomics;Whole slide images;.

\bibitem{zhou2024cohort}
H.~Zhou, F.~Zhou, and H.~Chen, ``Cohort-individual cooperative learning for multimodal cancer survival analysis,'' {\em IEEE Transactions on Medical Imaging}, pp.~1--1, 2024.

\bibitem{wang2024histo}
Z.~Wang, Y.~Zhang, Y.~Xu, S.~Imoto, H.~Chen, and J.~Song, ``Histo-genomic knowledge distillation for cancer prognosis from histopathology whole slide images,'' {\em arXiv preprint arXiv:2403.10040}, 2024.

\bibitem{li2023survival}
Z.~Li, Y.~Jiang, M.~Lu, R.~Li, and Y.~Xia, ``Survival prediction via hierarchical multimodal co-attention transformer: A computational histology-radiology solution,'' {\em IEEE Transactions on Medical Imaging}, vol.~42, no.~9, pp.~2678--2689, 2023.

\bibitem{jeong2024artificial}
D.~Y. Jeong, J.~Park, H.~Song, J.~Moon, T.~Lee, C.~Ahn, S.~Park, S.-H. Lee, C.-Y. Ock, and H.~Y. Lee, ``Artificial intelligence (ai)-based multi-modal approach using h\&e and ct image for predicting treatment response of immune checkpoint inhibitor (ici) in non-small cell lung cancer (nsclc),'' {\em Cancer Research}, vol.~84, no.~6\_Supplement, pp.~4170--4170, 2024.

\bibitem{liberzon2015molecular}
A.~Liberzon, C.~Birger, H.~Thorvaldsd{\'o}ttir, M.~Ghandi, J.~P. Mesirov, and P.~Tamayo, ``The molecular signatures database hallmark gene set collection,'' {\em Cell systems}, vol.~1, no.~6, pp.~417--425, 2015.

\bibitem{genai2023llama}
M.~GenAI, ``Llama 2: Open foundation and fine-tuned chat models,'' {\em arXiv preprint arXiv:2307.09288}, 2023.

\bibitem{li2022blip}
J.~Li, D.~Li, C.~Xiong, and S.~Hoi, ``Blip: Bootstrapping language-image pre-training for unified vision-language understanding and generation,'' in {\em International conference on machine learning}, pp.~12888--12900, PMLR, 2022.

\bibitem{gao2024clip}
P.~Gao, S.~Geng, R.~Zhang, T.~Ma, R.~Fang, Y.~Zhang, H.~Li, and Y.~Qiao, ``Clip-adapter: Better vision-language models with feature adapters,'' {\em International Journal of Computer Vision}, vol.~132, no.~2, pp.~581--595, 2024.

\bibitem{he2016deep}
K.~He, X.~Zhang, S.~Ren, and J.~Sun, ``Deep residual learning for image recognition,'' in {\em Proceedings of the IEEE conference on computer vision and pattern recognition}, pp.~770--778, 2016.

\bibitem{kantorovich2006translocation}
L.~V. Kantorovich, ``On the translocation of masses.,'' {\em Journal of mathematical sciences}, vol.~133, no.~4, 2006.

\bibitem{chen2022pan}
R.~J. Chen, M.~Y. Lu, D.~F. Williamson, T.~Y. Chen, J.~Lipkova, Z.~Noor, M.~Shaban, M.~Shady, M.~Williams, B.~Joo, {\em et~al.}, ``Pan-cancer integrative histology-genomic analysis via multimodal deep learning,'' {\em Cancer Cell}, vol.~40, no.~8, pp.~865--878, 2022.

\bibitem{selvaraju2017grad}
R.~R. Selvaraju, M.~Cogswell, A.~Das, R.~Vedantam, D.~Parikh, and D.~Batra, ``Grad-cam: Visual explanations from deep networks via gradient-based localization,'' in {\em Proceedings of the IEEE international conference on computer vision}, pp.~618--626, 2017.

\end{thebibliography}

\end{document}